\documentclass[aps,prl,reprint,superscriptaddress]{revtex4-1}
\usepackage{threeparttable}
\usepackage[dvips]{graphics}
\usepackage{longtable}
\usepackage{enumerate}
\usepackage{amsfonts}
\usepackage{amsmath}
\usepackage{stmaryrd}
\usepackage{dcolumn}
\usepackage{xcolor}
\usepackage{bm}
\usepackage{graphicx}
\begin{document}
\title{Accurate prediction and measurement of vibronic branching ratios for laser cooling linear polyatomic molecules}

\author{Chaoqun Zhang}
\affiliation{Department of Chemistry, The Johns Hopkins University, Baltimore, MD 21218, USA}

\author{Benjamin L. Augenbraun}
\email{augenbraun@g.harvard.edu}
\author{Zack D. Lasner}
\author{Nathaniel B. Vilas}
\author{John M. Doyle}
\affiliation{Department of Physics, Harvard University, Cambridge, MA 02138, USA}
\affiliation{Harvard-MIT Center for Ultracold Atoms, Cambridge, MA 02138, USA}

\author{Lan Cheng}
\email{lcheng24@jhu.edu}
\affiliation{Department of Chemistry, The Johns Hopkins University, Baltimore, MD 21218, USA}

\begin{abstract}

   We report 
   a generally applicable computational and experimental approach 
   to determine vibronic branching ratios
   in linear polyatomic molecules
   to the 10$^{-5}$ level, including for nominally symmetry-forbidden
   transitions.
   These methods are demonstrated in CaOH and YbOH,
   showing approximately two orders of magnitude {improved sensitivity} compared with the previous state of the art.
   Knowledge of branching ratios at this level is needed for the successful deep laser cooling of a broad range of molecular species.

\end{abstract}

\maketitle

  %introduction
  
  \textit{Introduction.}---Recent years have seen rapid progress in producing cold molecules and in using them 
  to study molecular interactions and reactions \cite{Ospelkaus2010, Quemener2012,bohn2017cold}, quantum information processing and computation \cite{demille2002quantum,carr2009cold,wei2011entanglement,karra2016prospects,hudson2018dipolar,ni2018dipolar,yu2019scalable}, 
  and precision measurement in search of new physics beyond the Standard Model (BSM) \cite{zelevinsky2008precision,kozyryev2017precision,YbF2018,Augenbraun20a}. 
  Direct laser cooling and trapping has emerged as a particularly promising approach \cite{di2004laser,Stuhl2008,carr2009cold,Shuman2010,%zeppenfeld2012sisyphus,
  Hummon13,zhelyazkova2014laser,Barry14,Kozyryev2016,SrOH2017,Truppe17,Anderegg17,Williams18,Collopy2018,baum20201d,Ding2020,mitra2020direct,augenbraun2020molecular} 
  to make molecules ultracold, as needed for much of the current and future research in these areas. 
  While molecules, compared to atoms, have wider applicability and 
  often have higher sensitivity for BSM searches, 
  their complex internal motions continue to pose challenges to applying the most powerful laser-cooling techniques. 
  {The use of magneto-optical traps (MOTs) and the very high-fidelity optical readout of internal quantum states typically rely on scattering more than} $10^4$ photons \cite{di2004laser,Stuhl2008};
  {to achieve this, a transition manifold with overall fractional loss} less than $10^{-4}$ is required.
  
  Several diatomic molecules have been optically cycled well above the required 10$^4$ photon threshold, leading to successful laser cooling \cite{Shuman2010,Hummon13} 
  and loading into 
  optical traps \cite{Barry14,Truppe17,Anderegg17,Williams18,Collopy2018}. 
  Laser cooling of polyatomic molecules, however, is more challenging.
  To understand the difficulty, this key metric should be considered - the quasidiagonality of
  the Franck-Condon matrix for the electronic laser cooling transition
  \cite{Isaev2016,Li2019,Hao2019,Ivanov2019,ivanov2020search,Dickerson21}
  In order to quantify this metric, and thus determine whether a molecular candidate can be successfully loaded and held in a MOT, knowledge of all {decays with intensity} above about $10^{-5}$ is needed {to ensure that a sufficient number of photons may be scattered by each molecule}. Recent observations of nominally symmetry-forbidden decays 
   ~\cite{baum2020establishing, Augenbraun2021YbOCH3} confirm that very weak decays, which can only be observed with considerable effort, can pose major obstacles to achieving an optical cycle capable of scattering $>10^4$ photons per molecule. While experimental studies using high resolution laser spectroscopy  \cite{Sheridan2007,Nakhate2019,Kozyryev2019,baum2020establishing,Nguyen2018,Mengesha2020,augenbraun2020molecular} 
  have the potential to provide accurate branching ratios, it is a formidable endeavor to measure branching ratios smaller than $10^{-3}$ with current methods. The capability to calculate and measure vibrational branching ratios accurately is thus 
  a critical tool for 
  developing laser-cooling strategies for specific molecules.  

  In this Letter, we present generally applicable methods to calculate and measure %method capable of calculating
  vibronic branching ratios for all transitions with intensity above $10^{-5}$ in laser-coolable linear polyatomic molecules. 
  Spin-vibronic perturbations \cite{Fieldbook} contribute significantly to vibronic decay pathways, so that modeling their effects on laser cooling is of utmost importance.
  The present computational scheme performs coupled-cluster (CC) \cite{Crawford00,Bartlett07,Stanton1993, Krylov08} calculations to parametrize a K{\"o}ppel-Domcke-Cederbaum (KDC) multi-state quasidiabatic Hamiltonian \cite{Koeppel84}  
  including all relevant spin-vibronic perturbations
  and obtains vibronic levels and wavefunctions from discrete variable representation (DVR) \cite{light1985generalized,Colbert1992,light2000discrete} calculations.  
  {We compare the computational results to new measurements of branching ratios in CaOH and YbOH, with approximately $10^{-5}$ relative intensity sensitivity. {These measurements, which rely crucially on strong optical excitation, achieve a level of sensitivity with direct experimental access to the perturbations probed by our calculations.}} {The {good agreement between experiment and theory demonstrated here validates both methodologies} for guiding the selection of candidate molecules for direct laser cooling {and for understanding the role that weak perturbations play in the optical cycling process.}}
  It also provides clear pathways for deep laser cooling of the two studied species, for use in quantum computation and the search of physics BSM. These methods in general provide an approach for assessing and understanding laser cooling of polyatomic molecules.

\begin{figure}
    \centering
    \includegraphics[width=8.6cm]{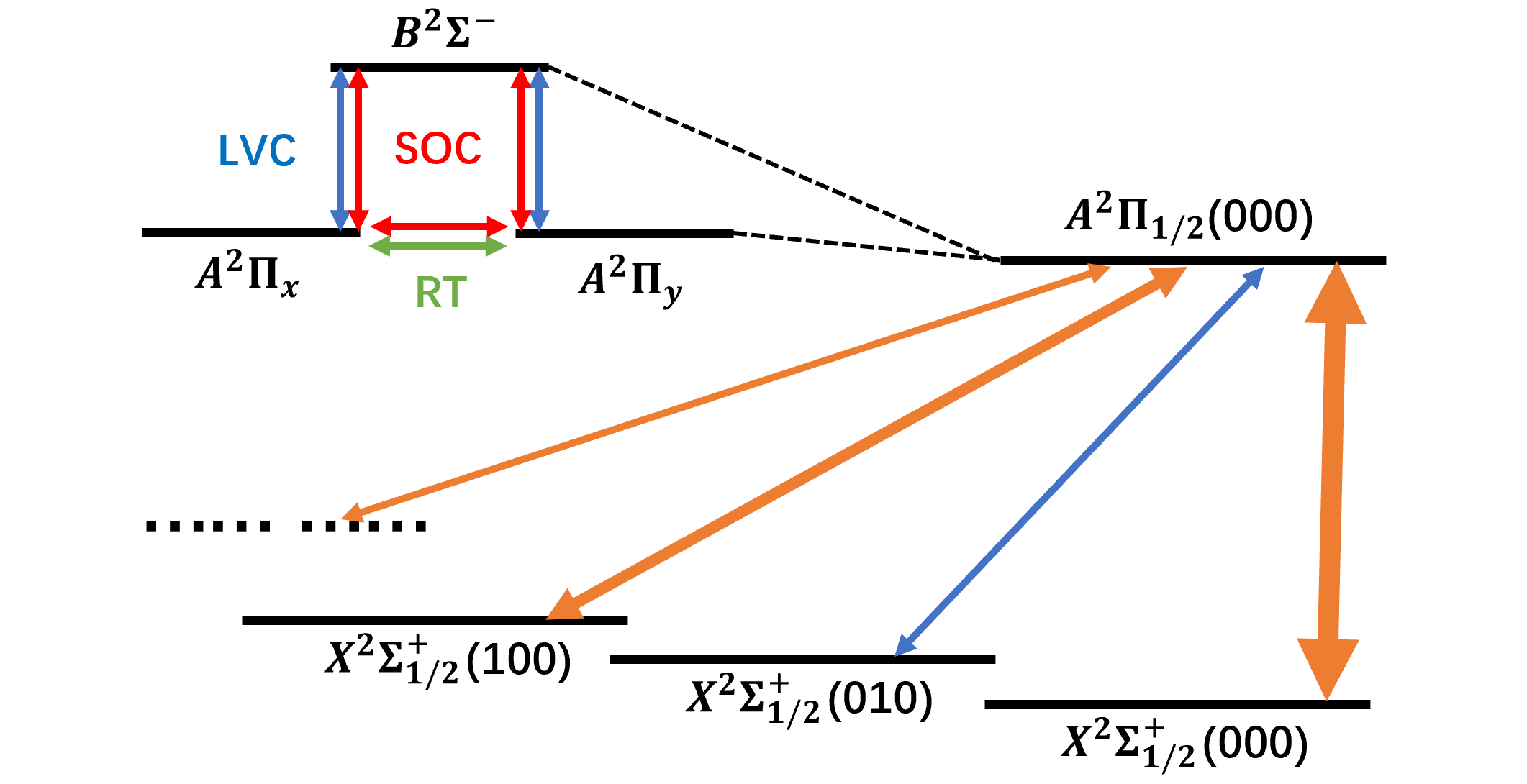}
    \caption{Schematic illustration of small perturbations contributing to the $A^2\Pi_{1/2}$(000) vibronic wavefunction
    as well as transitions to vibrational states of ground electronic states pertinent to laser cooling.
    SOC, LVC, and RT stand for spin-orbit coupling, linear vibronic coupling and Renner-Teller coupling, respectively. 
    The blue arrow denotes a nominally symmetry-forbidden transition. }
    \label{energy structure fig}
     \vspace{-10pt}
  \end{figure}

  \textit{Theory.}---
  As shown in Figure \ref{energy structure fig}, the most commonly used optical cycle in a linear triatomic molecule of the type M-A-B, e.g., a metal hydroxide, % or isocyanide, 
  involves the transitions from $A^2\Pi_{1/2}(000)$, the vibrational ground state of a low-lying excited electronic state, to the vibrational states of the ground electronic state $X^2\Sigma_{1/2}$, i.e., $X^2\Sigma_{1/2}(000)$, $X^2\Sigma_{1/2}(100)$, $X^2\Sigma_{1/2}(010)$, etc. Here %(n m l) 
  $(v_1 v_2^\ell v_3)$ denotes a vibrational state with $v_1$ quanta of excitation in the M-A stretching mode, $v_2$ quanta of bending excitation, and $v_3$ quanta of A-B stretching excitation. The superscript $\ell$ denotes a vibrational angular momentum due to bending excitations. $X^2\Sigma_{1/2}$ is well separated from the other electronic states. A variational vibrational calculation within the Born-Oppenheimer approximation thus can produce its vibrational levels and wave functions accurately. 
  
  In contrast, a number of small perturbations \cite{Fieldbook} contribute to the $A^2\Pi_{1/2}(000)$ vibronic wave function and hence to the targeted branching ratios. The Renner-Teller (RT) effects \cite{renner1934theorie} and spin-orbit coupling (SOC) are well known to play important roles in the description of vibronic levels and wave functions for the $A^2\Pi$ states in a linear triatomic molecule \cite{carter1984variational,carter1990}. Besides, importantly, linear vibronic coupling (LVC) and SOC between $A^2\Pi$ and $B^2\Sigma$ introduce $A^2\Pi$(010) and $B^2\Sigma$(010) components into the wave function of $A^2\Pi_{1/2}$(000) and thus make important contributions to nominally symmetry-forbidden transitions to $X^2\Sigma_{1/2}(010)$ \cite{Brazier1985,coxon1994laser} with small but non-negligible branching factors. Therefore, aiming to obtain accurate branching factors higher than $10^{-5}$, it is necessary to calculate the $A^2\Pi_{1/2}(000)$ vibronic wave function using a multi-state Hamiltonian with all of these relevant perturbations taken into account. 
  We thus adopt the %K{\"o}ppel-Domcke-Cederbaum (KDC) 
  KDC Hamiltonian %\cite{Koeppel84}
  widely used to account for spin-vibronic coupling in spectrum simulation
  \cite{Koppel81, SchmidtKlugmann03, Marenich04, Mishra06, Mishra07, Schuurman07, Garand10, Nagesh12, Shao13, Weichman17,Scheit19}.
  
  We use a KDC Hamiltonian of a form
  \begin{eqnarray}
  \begin{tiny}
      \left[ \begin{array}{cccccc}
          {E}^{\text{qd}}_{{A^2\Pi_x}}    & V^{\text{RT}}_{\text{xy}}+ih^{\text{SO}}_{\text{AA}} & V^{\text{LVC}}_{\text{xB}} & 0 & 0 & h^{\text{SO}}_{\text{AB}}\\ \\
          V^{\text{RT}}_{\text{yx}}-ih^{\text{SO}}_{\text{AA}} & {E}^{\text{qd}}_{{A^2\Pi_y}} & V^{\text{LVC}}_{\text{yB}} & 0 & 0 & ih^{\text{SO}}_{\text{AB}}\\ \\
          V^{\text{LVC}}_{\text{Bx}} & V^{\text{LVC}}_{\text{By}} & {E}^{\text{qd}}_{{B^2\Sigma}}  & -h^{\text{SO}}_{\text{BA}} & -ih^{\text{SO}}_{\text{BA}}& 0 \\ \\
           0 & 0  & -h^{\text{SO}}_{\text{AB}} & {E}^{\text{qd}}_{A^2\Pi_x} & V^{\text{RT}}_{\text{xy}}-ih^{\text{SO}}_{\text{AA}} & V^{\text{LVC}}_{\text{xB}} \\ \\
           0 & 0  & ih^{\text{SO}}_{\text{AB}} & V^{\text{RT}}_{\text{yx}}+ih^{\text{SO}}_{\text{AA}} & {E}^{\text{qd}}_{A^2\Pi_y}  & V^{\text{LVC}}_{\text{yB}} \\ \\
          h^{\text{SO}}_{\text{BA}}   & -ih^{\text{SO}}_{\text{BA}}  & 0 & V^{\text{LVC}}_{\text{Bx}} & V^{\text{LVC}}_{\text{By}}  & {E}^{\text{qd}}_{B^2\Sigma}     
      \end{array} \right]
  \end{tiny}
  \label{6sH}
  \end{eqnarray}
  including six scalar quasidiabatic electronic states, $A^2\Pi_x(\alpha)$, $A^2\Pi_y(\alpha)$, $B^2\Sigma(\alpha)$, 
  $A^2\Pi_x(\beta)$, $A^2\Pi_y(\beta)$, and $B^2\Sigma(\beta)$, as the electronic basis. 
  Here $\alpha$ and $\beta$ refer to $S_z = 1/2$ and $-1/2$. % for the unpaired electron in these electronic states. 
  ${E}^{\text{qd}}_{A^2\Pi_x}$, ${E}^{\text{qd}}_{A^2\Pi_y}$, and ${E}^{\text{qd}}_{B^2\Sigma}$ 
  denote potential energies of these quasidiabatic states.
  $V^{\text{RT}}$'s represent RT coupling, % between $A^2\Pi_x$ and $A^2\Pi_y$, 
  $V^{\text{LVC}}$'s represent LVC, % $V^{\text{LVC}}_{\text{xB}}$ and $V^{\text{LVC}}_{\text{yB}}$ represent LVC between $A^2\Pi$ and $B^2\Sigma$, 
  and $h^{\text{SO}}$'s %and $\bar{h}^{\text{SO}}$ 
  denote SOC matrix elements. This Hamiltonian includes all perturbations
  pertinent to the calculations of branching ratios for %pertinent to 
  the $A^2\Pi_{1/2}\to X^2\Sigma_{1/2}$ optical cycle.

  We first calculate the adiabatic potential energy surfaces (PESs) for $A^2\Pi_+$ and $A^2\Pi_-$, the two scalar adiabatic states of $A^2\Pi$, 
  as well as for $B^2\Sigma$,
  \begin{eqnarray}
      \left[ \begin{array}{ccc}
          E^{\text{ad}}_{A^2\Pi_+}    & 0 & 0 \\
          0 & E^{\text{ad}}_{A^2\Pi_-}    & 0 \\
          0 & 0 & E^{\text{ad}}_{B^2\Sigma}\\
      \end{array} \right].
  \label{3saH}
  \end{eqnarray}
  %Figure \ref{adt fig} provides a schematic illustration for
  %the adiabatic potentials energy surfaces of the $A^2\Pi$ states. 
  Then we apply the following adiabatic to diabatic transformation %(ADT) 
  \begin{equation}
      \left[\begin{array}{cc}
          \frac{Q_x}{\sqrt{Q_x^2+Q_y^2}} & \frac{Q_y}{\sqrt{Q_x^2+Q_y^2}} \\
          \frac{Q_y}{\sqrt{Q_x^2+Q_y^2}} & -\frac{Q_x}{\sqrt{Q_x^2+Q_y^2}}
      \end{array}\right],
      \label{ADT}
  \end{equation}
  in which $Q_x$ and $Q_y$ represent the normal coordinates of the two bending modes. 
  This transformation was derived for harmonic potentials. 
  The application to potentials with anharmonic contributions
  corresponds to a quasidiabatization process. 
  This transforms %the adiabatic 
  $A^2\Pi_+$ and $A^2\Pi_-$ %states 
  into two quasidiabatic states, 
  hereafter referred to as $A^2\Pi_x$ and $A^2\Pi_y$,
  and the adiabatic potentials in Eq. (\ref{3saH}) into the quasidiabatic potentials
   \begin{eqnarray}
         \left[ \begin{array}{ccc}
          \widetilde{E}^{\text{qd}}_{A^2\Pi_x}    & \widetilde{V}^{\text{RT}}_{\text{xy}} & 0 \\
          \widetilde{V}^{\text{RT}}_{\text{xy}} & \widetilde{E}^{\text{qd}}_{A^2\Pi_y}    & 0 \\
          0 & 0 & E^{\text{ad}}_{B^2\Sigma}\\
      \end{array} \right]   
  \label{3sqdH}
  \end{eqnarray}
%   A schematic representation
%   for both the adiabatic and quasidiabatic potential energy surfaces are given in the supplementary materials.
 
  Next, we include LVC using the linear diabatic coupling constants \cite{Ichino09} between $A^2\Pi_{x(y)}$ and $B^2\Sigma$.
  %\begin{eqnarray}
  %V^{\text{LVC}}_{\text{xB}}=\left. Q_x \frac{\partial \langle %A^2\Pi_x|H|B^2\Sigma\rangle}{\partial Q_x}\right|_{Q_x = 0}.
  %\end{eqnarray}
  This also introduces corrections to $\widetilde{E}^{\text{qd}}$, $\widetilde{V}^{\text{RT}}$, and $E^{\text{ad}}_{B^2\Sigma}$ \cite{Ichino09},
  leading to 
     \begin{eqnarray}
         \left[ \begin{array}{ccc}
          {E}^{\text{qd}}_{A^2\Pi_x}    & {V}^{\text{RT}}_{\text{xy}} & V^{\text{LVC}}_{\text{xB}} \\
          {V}^{\text{RT}}_{\text{xy}} & {E}^{\text{qd}}_{A^2\Pi_y}    & V^{\text{LVC}}_{\text{yB}} \\
          V^{\text{LVC}}_{\text{Bx}} & V^{\text{LVC}}_{\text{By}} & E^{\text{qd}}_{B^2\Sigma}\\
      \end{array} \right].  
  \label{3sqdH3}
  \end{eqnarray}
  Finally, we augment Eq. (\ref{3sqdH3}) with SOC and obtain
  Eq. (\ref{6sH}), hereby completing the construction of the quasidiabatic Hamiltonian. 
  
  We used the CFOUR program 
  \cite{cfour,Matthews2020a,Stanton1993,Stanton99,Cheng11b}
  to calculate all parameters in the KDC Hamiltonian
  including adiabatic PESs, linear diabatic coupling constants \cite{Ichino09}, and SOC matrix elements \cite{Klein08,cheng18a,zhang2020performance}.
  %Since laser cooling involves low-lying vibrational states, 
  %CC \cite{Crawford00,Bartlett07} 
  %and equation-of-motion (EOM) CC \cite{Stanton1993, Krylov08} methods that can accurately calculate the wave functions around the equilibrium structures are the methods of choice for these calculations.
  The present study used the EOM electron attachment 
  CC singles and doubles (EOMEA-CCSD) method \cite{Nooijen95}
  and correlation-consistent basis sets \cite{dunning1989gaussian,koput2002ab,de2001parallel,hill2017gaussian,lu2016correlation} to provide accurate PESs around the equilibrium structures pertinent
  to the calculations of low-lying vibronic states involved
  in laser cooling.
  We accounted for scalar-relativistic effects for SrOH and YbOH 
  using the spin-free exact two-component theory in its one-electron variant (SFX2C-1e) 
  \cite{Dyall2001, Liu2009}. 
  %More details are documented in the Supporting Information.
   Having the quasidiabatic Hamiltonians expanded in ``real space" basis sets, %. One advantage of using normal coordinates DVR is that the calculations can naturally reproduce the splittings between excited bending modes like the splitting between $(02^00)$ and $(02^20)$ statesWith the availability of the hamiltonian parameters, 
  we carried out DVR \cite{Colbert1992} calculations in the representation
  of the four vibrational normal coordinates
  to obtain vibronic levels and wave functions.
  The normal coordinate DVR calculations previously produced accurate vibrational levels
  for the $X^2\Sigma$ state of YbOH \cite{Mengesha2020}. 
  %including the splittings between $(02^00)$ and $(02^20)$.
  We then combine the Franck-Condon overlap integrals with 
  the EOMEA-CCSD electronic transition dipole moments to 
 obtain the branching factors. 
  We refer the readers to the supporting information \cite{SI} for 
  more details about the computations. 

  \begin{table}
  \begin{center}
  \caption{Relative vibrational energy levels (in cm$^{-1}$) of the $X^2\Sigma_{1/2}$ and $A^2\Pi_{1/2}$ states in CaOH and SrOH. }
   \vspace{-10pt}
  \label{vib ene table}
  \begin{tabular}{ccccccccc}
    \hline \hline
    ~& \multicolumn{4}{c}{CaOH} &\multicolumn{4}{c}{SrOH}\\
    States & ~& Calculated & ~&Exp.$^a$ & ~& Calculated & ~&Exp. \cite{Presunka1995,Nguyen2018}\\
    \hline
    $X^2\Sigma_{1/2}(000)$ &~& 0 &~& 0.0 &~& 0 &~& 0.0 \\
    $X^2\Sigma_{1/2}(010)$ &~& 355 &~& 352.9 &~& 367  &~& 363.7\\ 
    $X^2\Sigma_{1/2}(100)$ &~& 611 &~& 609.0 &~& 534  &~& 527.0\\
    $X^2\Sigma_{1/2}(02^00)$ &~& 695 &~& 688.7 &~& 702 &~ & 703.3\\
    $X^2\Sigma_{1/2}(02^20)$ &~& 718 &~& 713.0 &~& 736 &~& 733.5\\
    $X^2\Sigma_{1/2}(110)$ &~& 954 &~& 952 &~& 892 &~& -- \\
    $X^2\Sigma_{1/2}(200)$ &~& 1214 &~& 1210.2 &~& 1063 &~& 1049.1\\
    $A^2\Pi_{1/2}(000)$ &~& 0 &~& 0.0 &~& 0 &~ & 0.0    \\
    $A^2\Pi_{3/2}(000)$ &~& 67 &~& 66.8 &~& 254 &~ & 263.5    \\
    $A^2\Pi_{1/2}(010)$ &~& 346 &~& 345 &~& 379 &~ & 377.8       \\
    $A^2\Pi_{1/2}(010)$ &~& 362 &~& 360 &~& 383 &~ & 380.4 \\
    $A^2\Pi_{1/2}(100)$ &~& 623 &~& 628.7 &~& 552 &~ & 542.1\\
    $A^2\Pi_{3/2}(010)$ &~& 428 &~& 425 &~& 636 &~ & 641.8 \\
    $A^2\Pi_{3/2}(010)$ &~& 446 &~& 445 &~& 642 &~ & 648.4 \\
    $A^2\Pi_{3/2}(100)$ &~& 690 &~& 695.9 &~& 808    &~ & 806.6 \\
    \hline \hline
  \end{tabular}
  \end{center}
     \vspace{-5pt}
      a. Levels for the $X^2\Sigma$ states from Refs. \cite{Coxon1992,Li1995,Pereira1996}. Levels for the $A^2\Pi$ states from Refs. \cite{coxon1994laser,Li1995,Li1996}. $A^2\Pi(010)$ levels taken as $J=0$ results in Figure 3 of Ref. \cite{Li1995}.
   %   b. From Ref. \cite{Presunka1995,Nguyen2018}
   \vspace{-15pt}
\end{table}

\textit{Experiment.}--- Branching ratios for CaOH and YbOH were recorded using dispersed laser-induced fluorescence (DLIF) measurements.  The sensitivity achieved was improved by approximately two orders of magnitude over previous measurements in similar species \cite{Kozyryev2019,Paul2019,Nguyen2018, Mengesha2020}, due primarily to the use of bright cryogenic molecular beams and photon cycling described below and in \cite{SI}.
  %This allowed us to achieve nearly a factor of 100 improvement over previous branching ratio measurements in similar species~\cite{Kozyryev2019,Paul2019,Nguyen2018, Mengesha2020}.
%  

In brief, cryogenic buffer gas beams of CaOH and YbOH were produced using the beam source described in Ref.~\cite{Augenbraun20a}. Laser-excitation-enhanced chemical reactions between metallic Yb or Ca and H$_2$O vapor~\cite{Fernando1991, Bopegedera1987alkylamide, Jadbabaie2020} increased the molecular beam flux by a factor of $\sim$10 compared to previous reports~\cite{baum20201d, Augenbraun20a}. The apparatus was modified with a spectrometer similar to that used in Ref.~\cite{Augenbraun2021YbOCH3}. After propagating approximately 30~cm downstream from the source, the molecules interacted with laser beams addressing the rotationally resolved $\tilde{A}\,^2\Pi_{1/2}(000)\leftarrow\tilde{X}\,^2\Sigma^+(000)$ and $\tilde{A}\,^2\Pi_{1/2}(000)\leftarrow\tilde{X}\,^2\Sigma^+(100)$ transitions used for optical cycling in laser cooling experiments~\cite{baum20201d, Augenbraun20a}. In this detection region, each molecule scattered on average 50-100 photons. Using optical cycling transitions therefore directly increased the number of photons collected, and thus the sensitivity, compared to typical DLIF measurements where molecules scatter on average only a single photon~\cite{Kozyryev2019}. 

    \begin{figure}
    \centering
    \includegraphics[width=8cm]{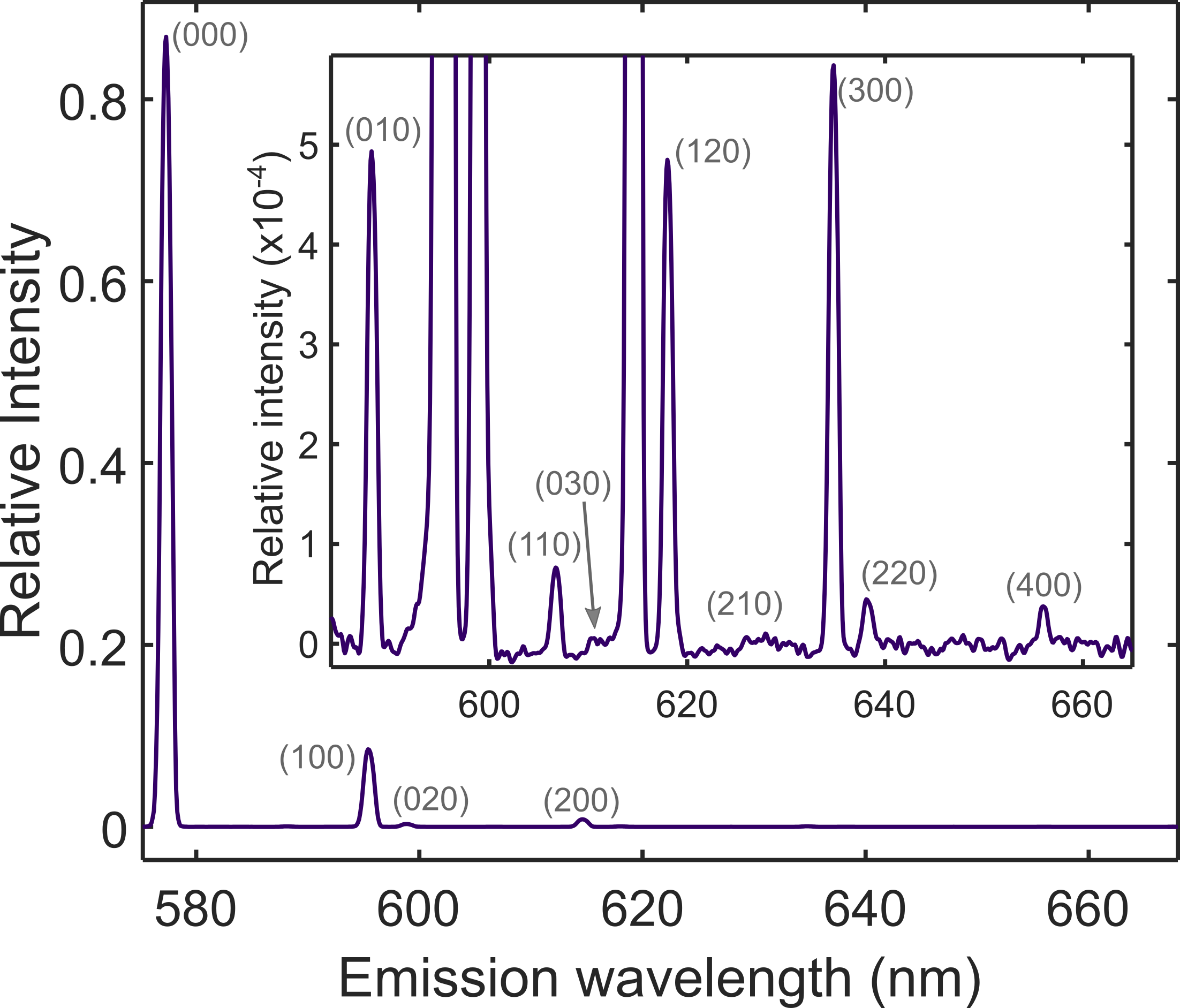}
    \caption{Dispersed laser-induced fluorescence measurement following excitation of YbOH to the $\tilde{A} \, ^2\Pi_{1/2}(000)$ $(J^\prime=1/2, p=+)$ state. Gray labels above each peak identify the ground state vibrational level $(v_1,v_2,v_3)$ populated by that decay. Inset: additional detail near the noise floor, demonstrating that spectral features with relative intensities as low as about $2 \times 10^{-5}$ can be identified.
    }
    \label{fig:DLIFdata}
       \vspace{-10pt}
  \end{figure}

The molecular fluorescence was collected and dispersed on a 0.67~m Czerny-Turner style spectrometer, before being detected with an electron-multiplying charged-coupled device (EMCCD). The wavelength and intensity axes were carefully calibrated in separate measurements (see \cite{SI}). See the supporting information \cite{SI} for more details on the experimental apparatus, calibration, and data analysis. A representative DLIF spectrum recorded for YbOH is shown in Fig. \ref{fig:DLIFdata}, demonstrating the sensitivity achieved by this method.

  \textit{Results and discussion.}---The computed vibronic energy levels for SrOH and CaOH, as summarized in Table \ref{vib ene table}, show
  excellent agreement with measured values, 
  with discrepancies  lower than 10 cm$^{-1}$.  The calculations also accurately reproduced
  the splittings between $(02^20)$ and $(02^00)$ and among the four $A^2\Pi(010)$ states.
  The present KDC Hamiltonian calculations thus accurately describe the vibronic wave functions. 
  We provide a complete list of computed vibronic levels for 
  the $X^2\Sigma$, $A^2\Pi$, and $B^2\Sigma$ states in the Supporting Information \cite{SI}. 
  
   \begin{table}
  \begin{center}
  \caption{Branching ratios for transitions from $A^2\Pi_{1/2}(000)$ to vibrational levels of $X^2\Sigma_{1/2}$ in CaOH and SrOH. The noise level of the smallest branching ratios for CaOH is at the level of $7\times 10^{-6}$; the uncertainty in the larger branching ratios is dominated by spectrometer calibration. The relative intensities of the unresolved $\ell$$=$0, 2 components of the (020) and (120) manifolds were fixed to the value measured in previous work \cite{baum2020establishing}.}
  \label{br table A}
  \begin{tabular}{ccccccccc}
    \hline \hline
     & ~&\multicolumn{3}{c}{CaOH}&~& \multicolumn{2}{c}{SrOH} \\
    States & ~& Calculated & ~& Experiment &~&  Calculated & Experiment\cite{Nguyen2018}  \\
    \hline
    $(000)$     &~& 95.429\%   &~& 94.59(29)\%   &~& 94.509\%   & 95.7\%  \\
    $(010)$     &~& 0.063\%    &~& 0.099(6)\%    &~& 0.025\%    & --      \\
    $(100)$     &~& 3.934\%    &~& 4.75(27)\%    &~& 5.188\%    &  4.3\%  \\
    $(02^00)$   &~& 0.298\%    &~& 0.270(17)\%   &~& 0.010\%    & --      \\
    $(02^20)$   &~& 0.079\%    &~& 0.067(12)\%   &~& 0.036\%    & --      \\
    $(110)$     &~& 0.003\%    &~& 0.0064(7)\%   &~& 0.001\%    & --      \\
    $(03^10)$   &~& $<$0.001\% &~& 0.0034(8)\%   &~& $<$0.001\% & --      \\
    $(200)$     &~& 0.157\%    &~& 0.174(16)\%   &~& 0.218\%    & --      \\
    $(12^00)$   &~& 0.022\%    &~& 0.021(3)\%    &~&$<$0.001\%  & --      \\
    $(12^20)$   &~& 0.005\%    &~& 0.005(1)\%    &~& 0.003\%    & --      \\
    $(040)$     &~& $<$0.001\% &~& 0.0021(7)\%   &~& $<$0.001\% & --      \\
    $(300)$     &~& 0.006\%    &~& 0.0068(8)\%   &~& 0.008\%    & --      \\
    $(22^00)$   &~& 0.002\%    &~& 0.0020(7)\%   &~& $<$0.001\% & --      \\
    \hline \hline
  \end{tabular}
  \end{center}
  \vspace{-25pt}
\end{table}

%Tables \ref{br table A}, \ref{br table B}, and \ref{brtableC} summarize the branching ratios for the transitions from $A^2\Pi_{1/2}(000)$ to the vibrational levels of $X^2\Sigma$ in CaOH, SrOH, and YbOH that are greater than $10^{-5}$.
Tables \ref{br table A} and \ref{brtableC} summarize the calculated and measured vibrational branching ratios for decay from $A^2\Pi_{1/2}(000)$ to the $X^2\Sigma$ state in CaOH and YbOH. All branching ratios above $10^{-5}$ are shown; computed branching to the O-H stretch mode is below 10$^{-6}$ and therefore negligible.
The present level of consistency between computations and measurements for CaOH and YbOH,
i.e., excellent agreement for stronger transitions
and qualitative agreement for weaker transitions below 0.1\%,
is very promising.
In CaOH, although branching ratios for the origin transitions are as large as $\sim$95\%,
one has to take into account up to (300) to saturate the {Ca-O stretch} branching ratios to below $10^{-5}$.
In YbOH, the branching ratios for the Yb-O stretch are less diagonal and (400) is relevant. 

\begin{table}
  \begin{center}
  \caption{The vibrational level positions (in cm$^{-1}$) of the $X^2\Sigma$ state of YbOH and vibrational branching ratios (VBRs) of the transitions from $A^2\Pi_{1/2}(000)$ to these states. The measured vibrational frequencies have an estimated uncertainty of $\pm 5$~cm$^{-1}$ and the noise level of the smallest branching ratios is at the level of $9 \times 10^{-6}$. Anharmonic splittings in the $(020)$ and $(120)$ levels were not resolved. 
  }
  \label{brtableC}
  \begin{tabular}{cccccc}
    \hline \hline
                       & \multicolumn{2}{c}{Levels} & \multicolumn{2}{c}{VBRs}   \\
           States            & Calculated & Experiment & Calculated & Experiment \\
    \hline
    $(000)$   & 0     & 0   & 87.691\% & 89.44(61)\% \\
    $(010)$   & 322   & 319  & 0.053\% &  0.054(4)\% \\
    $(100)$   & 532   & 528  & 10.774\% & 9.11(55)\%\\
    $(02^00)$ & 631   & 627  & 0.457\% & 0.335(20)\%\\
    $(02^20)$ & 654   & --   & 0.022\% & --\\
    $(110)$   & 846   & 840  & 0.007\% & 0.0100(13)\% \\
    $(03^10)$   & 952  & 947 & $<$0.001\% & 0.0020(9)\% \\
    $(200)$   & 1062  & 1052 & 0.863\% & 0.914(62)\% \\
    $(12^00)$ & 1154  & 1144 & 0.063\% & 0.055(4)\% \\
    $(12^20)$ & 1173  & --   & 0.004\% & -- \\
    $(210)$   & 1369  & 1369 & $<$0.001\% & 0.0019(12)\% \\
    $(300)$   & 1600  & 1572 & 0.054\% & 0.067(4)\% \\
    $(22^00)$ & 1680  & 1651   & 0.007\% & 0.0050(9)\% \\
    $(400)$   & 2160  & 2079 & 0.002\% & 0.0045(9)\% \\
    \hline \hline
  \end{tabular}
  \end{center}
     \vspace{-25pt}
\end{table}

\begin{figure}
    \centering
    \includegraphics[width=8.6cm]{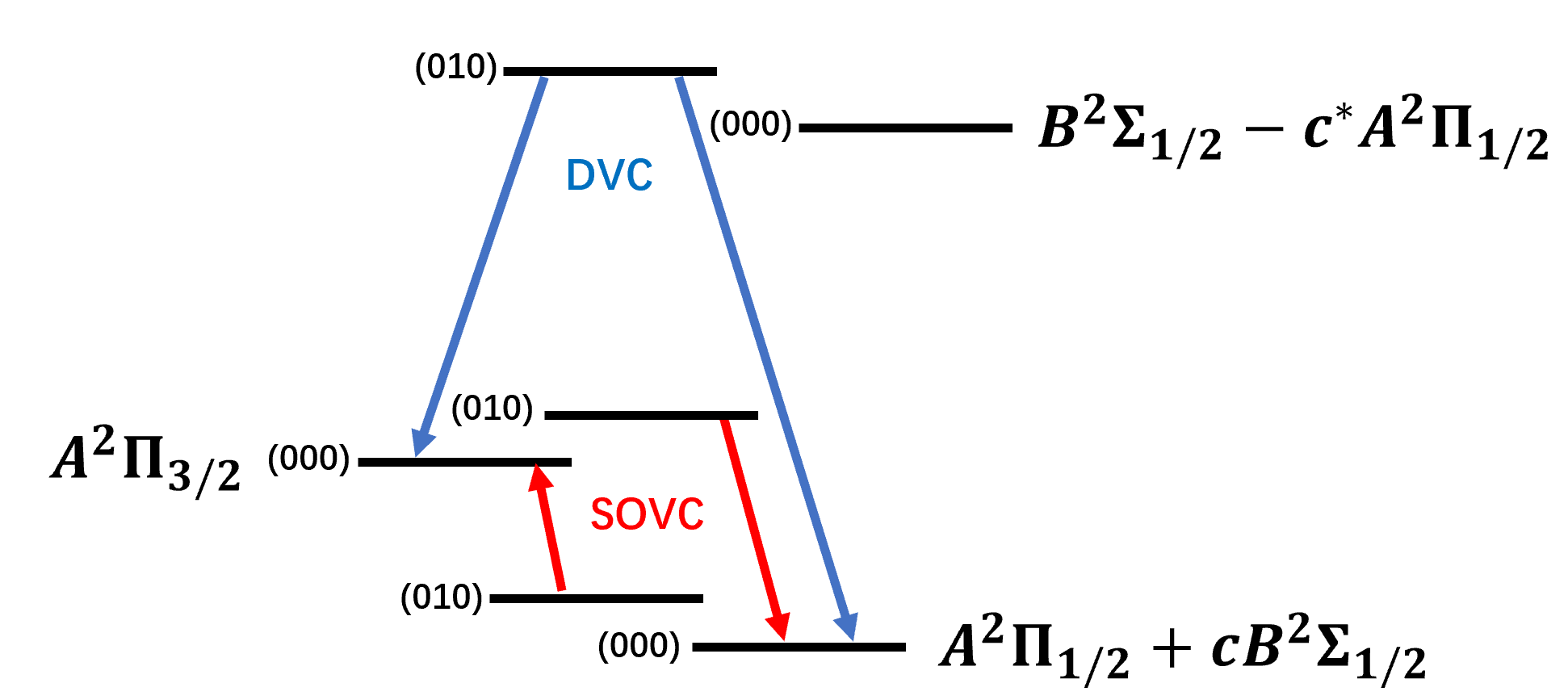}
    \caption{The direct vibronic coupling (DVC) and spin-obit-vibronic coupling (SOVC) mechanisms
    for borrowing intensities for the nominally symmetry-forbbiden $A^2\Pi(000)\rightarrow X^2\Sigma(010)$ transition. $c=\frac{h^{\text{SO}}_{\text{BA}}}{E_{\text{A}}-E_{\text{B}}}$ accounts for the SO mixing of A and B states. 
    }
    \label{mechfig}
       \vspace{-10pt}
  \end{figure}

  The nominally symmetry-forbidden $\left|\Delta v_2\right| = 1, 3, 5, \cdots$ transitions 
  borrow intensities through vibronic and/or spin-orbit coupling \cite{Brazier1985,coxon1994laser,baum2020establishing}.
  As shown in Figure \ref{mechfig}, %Take the $A^2\Pi(000)-X^2\Sigma(010)$ transition as an example. 
  the ``direct vibronic coupling" (DVC) mechanism %originates from vibronic coupling 
  mixes $B^2\Sigma(010)$ into the $A^2\Pi_{1/2}(000)$ wave function through LVC.
  The ``spin-orbit-vibronic coupling" (SOVC) mechanism 
  couples $B^2\Sigma(000)$ with $A^2\Pi(000)$ through SOC
  and then mixes %$A^2\Pi(010)$ with $B^2\Sigma(000)$ through linear vibronic coupling and hence introduces
  $A^2\Pi_{3/2}(010)$ into the $A^2\Pi_{1/2}(000)$ wave function through LVC. 
  A perturbative analysis gives the following ratio between the SOVC and DVC contributions
  to the intensity:
  \begin{eqnarray}
  \frac{I_{\text{SOVC}}}{I_{\text{DVC}}}=
  2\frac{|h^{\text{SO}}_{\text{BA}}|^2 |h^{\text{dip}}_{\text{AX}}|^2}{|\omega(010)+2h^{\text{SO}}_{\text{AA}}|^2 |h^{\text{dip}}_{\text{BX}}|^2}, \label{mech}
  \end{eqnarray}
  in which $h^{\text{dip}}_{\text{AX}}$ and $h^{\text{dip}}_{\text{BX}}$
  are the A-X and B-X electronic transition dipole moments. 
  %Using the $B^2\Sigma(010)$ and $A^2\Pi(010)$ components in the computed $A^2\Pi_{1/2}(000)$ wave function, we obtained 
  The percentages of the DVC contributions obtained using the $B^2\Sigma(010)$ and $A^2\Pi(010)$ components in the computed $A^2\Pi_{1/2}(000)$ wave function amount
  to 98\%, 89\%, and 70\% for CaOH, SrOH, and YbOH, respectively,
  which are consistent with the values of 99\%, 92\%, and 76\% obtained from the perturbative analysis
  using Eq. (\ref{mech}). % and computed parameters.  
  The DVC channel dominates this intensity borrowing process even 
  in molecules containing heavy atoms, since the effects of large $h^{\text{SO}}_{\text{BA}}$ 
  are offset by large $h^{\text{SO}}_{\text{AA}}$ in the denominator. % in Eq. \ref{mech}.

It is desirable to minimize the coupling between $X^2\Sigma$(100) and $X^2\Sigma$(020)
in designing laser-coolable linear triatomic molecules. 
As shown in Table \ref{br table A},
the intensities of the transitions to $X^2\Sigma$(020) in SrOH are significantly lower
than in CaOH.
This is due to the larger energy separation between the $X^2\Sigma$(100) and $X^2\Sigma$(020) levels in SrOH (534 cm$^{-1}$ and 702 cm$^{-1}$)
compared to CaOH (611 cm$^{-1}$ and 694 cm$^{-1}$), and hence weaker coupling.
{Similarly, there is non-negligible vibrational branching to $X^2\Sigma(120)$ and $X^2\Sigma(220)$ of CaOH; in contrast, the branching ratio to $X^2\Sigma(120)$ in SrOH is lower than $10^{-6}$.}

  \textit{Conclusion.}---{We have presented a computational and experimental scheme to determine vibronic branching ratios pertinent to scattering around 100,000 photons in linear polyatomic molecules.}
  {The computational methodology} is generally applicable to the study of optical cycles for linear triatomic molecules, such as metal hydroxides and isocyanides
  proposed for laser cooling. %being actively pursued in experiments. 
  {We have found good agreement with new experimental measurements that used optical cycling to achieve relative intensity sensitivities at the $10^{-5}$ level. The agreement supports the accuracy of both theoretical and experimental techniques.} 
  
  Further work for the future includes study of the accuracy %of the parameters in 
  for the parametrization of the quasidiabatic Hamiltonian, including the higher-order correlation effects on PESs and the effects of quadratic 
%$A^2\Pi-B^2\Sigma$ 
vibronic coupling,
%and the geometrical dependence of 
%SOC parameters,
 aiming to obtain quantitative accuracy for weak transitions.
  It is also of interest to study transitions involving
  dark electronic states \cite{Collopy2015,Collopy2018,Ding2020,Zhang2020yo} in linear triatomic molecules such as BaOH.  
    The generalization to calculations of linear tetraatomic molecules, e.g., metal monoacetylides, 
  should also provide very interesting results. 
  The calculations of nonlinear molecules is the next frontier,
  and we assess that we can follow the same generic scheme, 
  although the perturbations relevant to the calculations will be molecule specific. 
  The computational framework presented here thus forms a basis for calculations of vibronic levels and branching ratios that are sufficiently accurate and complete to facilitate and guide laser-cooling experiments for even highly complex polyatomic molecules.

 \textit{Acknowledgements.}---C. Z. thanks Yifan Shen (Baltimore) for stimulating discussions. 
 L. C. is grateful to Paul Dagdigian (Baltimore) and John Stanton (Gainesville) for reading the manuscript and for helpful comments.
  This computational work at the Johns Hopkins University is supported by 
  the National Science Foundation, under grant No. PHY-2011794.
  All calculations were carried out at the Maryland Advanced Research Computing Center (MARCC). The CaOH measurements were supported by the AFOSR and NSF. The YbOH measurements were supported by the Heising-Simons Foundation, the Gordon and Betty Moore Foundation, and the Alfred P. Sloan Foundation. N. B. V. acknowledges support from the NDSEG fellowship.

\bibliographystyle{apsrev4-1} % Tell bibtex which bibliography style to use
\bibliography{sroh} 

\clearpage

\end{document}

\title{Supplementary Information for ``Accurate prediction and measurement of vibronic branching ratios for laser cooling linear polyatomic molecules"}

\maketitle

\renewcommand{\theequation}{S.\arabic{equation}}
\renewcommand{\thetable}{S.\arabic{table}}

\section{Details of experimental measurements}
The vibrational frequencies and branching ratios for CaOH and YbOH, reported in Tabs. II and III of the main text, were measured using dispersed laser-induced fluorescence (DLIF) spectroscopy. A molecular beam of CaOH was produced using the cryogenic buffer-gas beam source described in Ref.~[\onlinecite{Augenbraun20a}]. The setup was slightly modified, as shown in Fig.~\ref{fig-Setup}. In brief, a target of solid Yb metal was ablated in the presence of $^4$He buffer gas at densities $\approx 10^{15}$~cm$^{-3}$ and temperatures around 2~K. Water vapor, introduced via a hot capillary held at 275~K, was allowed to react with the ablated Yb atoms and form, among other products, YbOH. Production of YbOH was enhanced by more than an order of magnitude by driving the Yb $^1S_0 \rightarrow$ $^3P_1$ intercombination line with approximately 1~W of laser power near 556~nm~[\onlinecite{Jadbabaie2020}]. The YbOH molecules were extracted through a 7~mm aperture into a beam and propagated approximately 40~cm to a room-temperature detection region. CaOH molecules were produced using identical methods, though using a Ca metal ablation target.

In the detection region, the molecules interacted with two laser beams driving the $\tilde{A}\,^2\Pi_{1/2}(000)\leftarrow\tilde{X}\,^2\Sigma^+(000)$ and $\tilde{A}\,^2\Pi_{1/2}(000)\leftarrow\tilde{X}\,^2\Sigma^+(100)$ transitions. Each laser beam carried two frequencies in order to drive the rotationally-closed $^pQ_{12}(1)$ and $P_{1}(1)$ lines. In this way, each molecule scattered on average $\sim150$ photons in the detection region. Note that because all laser excitation populated the same, single excited state ($J^\prime = 1/2,p=+$), the optical cycling does not introduce any artifacts. The cycling does, however, allow us to populate the excited state approximately a factor of 150$\times$ more often than would occur in an experiment that used only a single photon scatter.

    \begin{figure}
    \centering
    \includegraphics[width=0.75\columnwidth]{Experimental Setup.png}
    \caption{Schematic diagram (not to scale) showing the experimental setup for CaOH and YbOH branching ratio measurements. Depending on the molecular target, the ablation target was composed of either Yb or Ca metal. The collected laser-induced fluorescence is sent to a Czerny-Turner style spectrometer (not shown) and imaged onto an EMCCD. Inset: the vibrational levels addressed to achieve optical cycling in the detection region.}
    \label{fig-Setup}
  \end{figure}

The scattered light was collimated by a 35~mm diameter lens and sent to a 0.67~m Czerny-Turner style spectrometer (McPherson model 207A). A spherical mirror of 25~mm diameter opposite the collection lens increased the collection efficiency by about 30\%. The spectrometer's output was detected by a cooled electron-multiplying charge-coupled device (EMCCD; Andor iXon 897). Given the spectrometer's dispersion and the EMCCD's sensor, we were able to image a region of approximately 50~nm in a single dataset. Typically, around 1000 molecular beam pulses were probed before the spectrometer angle was stepped to image a new wavelength region.

The spectrometer output was calibrated as follows. The wavelength axis was calibrated by allowing a small amount of laser scatter at various wavelengths to pass through the spectrometer. The laser wavelength was read to MHz precision on a commercial wavemeter and mapped onto the imaged position on the EMCCD sensor. The relative intensity response of the spectrometer and EMCCD were simultaneously calibrated by coupling a few hundred nW of laser power at various wavelengths through the spectrometer and recording the number of counts detected as a function of wavelength. The laser power itself was calibrated using a commercial power meter. Finally, the calibration curve was corrected to normalize for constant photon flux (rather than constant power), and all data were normalized using this calibration.

    \begin{figure}
    \centering
    \includegraphics[width=0.85\columnwidth]{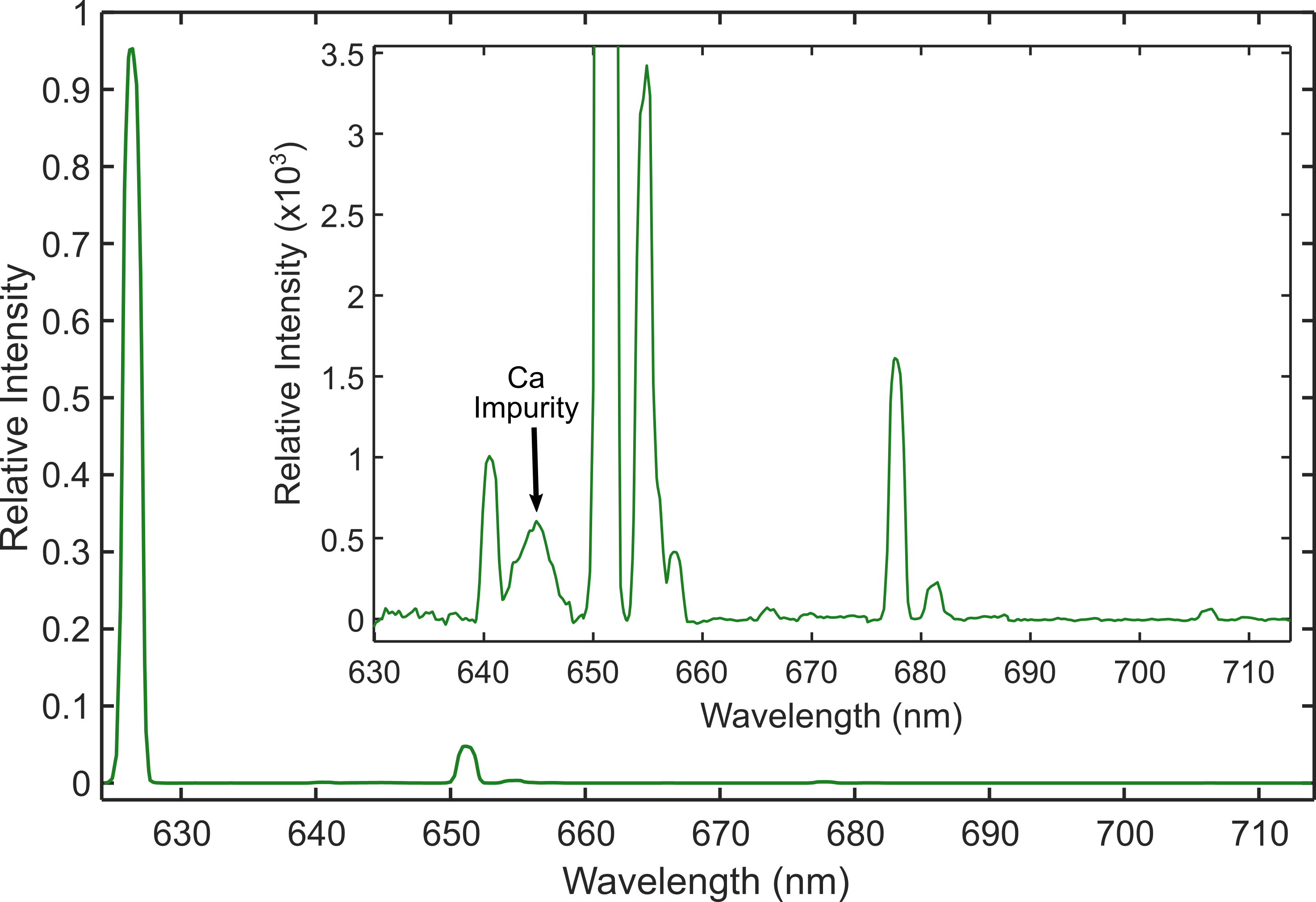}
    \caption{DLIF trace following excitation of CaOH to the $\tilde{A}\,^2\Pi_{1/2}(000)$ state. Inset: Enlarged region near the baseline of the relative intensity measurement demonstrating the noise level $<10^{-5}$ relative to the origin band.}
    \label{CaOH-data}
  \end{figure}

For most states, the dominant error in the branching ratios is from spectrometer calibration, which was limited by the $\sim$5\% calibration and readout uncertainty of the commercial power meter used. Additionally, the smallest observed branching ratios were limited by the noise floor of the data, which was $\sim$$7\times 10^{-6}$ for CaOH and $\sim$$9\times 10^{-6}$ for YbOH. Small uncertainties that may arise from data analysis are also included in the error bars, though these are typically small compared to the other two error sources.

We note that the DLIF measurements reported here for CaOH are consistent with recent measurements using a complementary approach based on optical pumping after cycling hundreds of photons [\onlinecite{baum2020establishing}]. Table \ref{tab:CaOHExpComparison} provides a direct comparison of these results.

  \begin{table}
  \begin{center}
  \caption{Comparison of the CaOH $A^2\Pi_{1/2}\rightarrow X^2\Sigma_{1/2}$  branching ratio measurements with results reported in Ref. [\onlinecite{baum2020establishing}]. Branching to the $N$$=$$1$ component of the (010) level is estimated by combining the rotationally-unresolved DLIF measurement with known \emph{relative} branching intensities to $N$$=$$1$ vs. $N$$=$$2$. These are empirically determined to be $\sim$0.65/0.35 for the $N$$=$1/2 rotational components from measurements similar to those in Ref. \cite{baum2020establishing}.}
  \label{tab:CaOHExpComparison}
  \begin{tabular}{r|rrrrrr}
    \hline \hline
    State & (000) & (100) & (200) & (02$^0$0) & (02$^2$0) & (010) \\
    & & & & & & $N$$=$$1$ \\
    \hline
    Ref. \cite{baum2020establishing} & 94.92(27)\% & 4.26(19)\% & 0.25(3)\% & 0.33(4)\% & 0.082(1)\% & 0.064(1)\% \\
    This work & 94.59(29)\% & 4.75(27)\% & 0.174(16)\% & 0.270(17)\% & 0.067(12)\% & 0.064(4)\% \\
    \hline \hline
  \end{tabular}
  \end{center}
\end{table}

\clearpage

\section{Details for the calculations of the parameters in the quasidiabatic Hamiltonian}

The quasidiabatic electronic energies and Renner-Teller coupling potential in Eq. (4) can be written as 
the linear combinations of adiabatic energies
\begin{equation}
  \label{after RT}
  \begin{aligned}
    & \widetilde{E}^{\text{qd}}_{A^2\Pi_x}=c_1^2 E^{\text{ad}}_{A^2\Pi_+}  + c_2^2 E^{\text{ad}}_{A^2\Pi_-},\\
    & \widetilde{E}^{\text{qd}}_{A^2\Pi_y}=c_1^2 E^{\text{ad}}_{A^2\Pi_-}  + c_2^2 E^{\text{ad}}_{A^2\Pi_+},\\
    & \widetilde{V}^{\text{RT}}_{\text{xy}}=c_1c_2\left[E^{\text{ad}}_{A^2\Pi_+} - E^{\text{ad}}_{A^2\Pi_-}\right],
  \end{aligned}
\end{equation}
where $c_1 = \frac{Q_x}{\sqrt{Q_x^2+Q_y^2}}$ and $c_2 = \frac{Q_y}{\sqrt{Q_x^2+Q_y^2}}$
with $Q_x$ and $Q_y$ representing normal coordinates for the bending modes. % are the transformation coefficients. Introduce 
The linear vibronic coupling is written as
\begin{eqnarray}
  V_{\text{Bp}}^{\text{LVC}}=V_{\text{pB}}^{\text{LVC}}=\lambda_p Q_p,
\end{eqnarray}
in which $p = x$ or $y$ and linear diabatic coupling constants were evaluated using the scheme
in Ref. [\onlinecite{Ichino09}].
Since the EOM-CCSD [\onlinecite{Stanton1993}] formulation is not Hermitian,
the matrix elements $\lambda_p$'s %$V^{\text{LVC}}_{\text{xB}}$ and $V^{\text{LVC}}_{\text{Bx}}$
are taken as the geometric average value of $\lambda^{\text{Bp}}_p$ and $\lambda^{\text{pB}}_p$
\begin{eqnarray}
  \lambda_p=sig(\lambda^{\text{Bp}}_p)\sqrt{\lambda^{\text{Bp}}_p \lambda^{\text{pB}}_p},
\end{eqnarray}
in which $\lambda^{\text{Bp}}_p$ and $\lambda^{\text{pB}}_p$ are given by
%between $A^2\Pi_{x(y)}$ and $B^2\Sigma$ states, which is evaluated using "fixed CI vector" scheme in EOM-CC equations, e.g.,
\begin{eqnarray}
  \lambda^{\text{Bp}}_p \equiv \left.  \frac{\partial \langle \Phi_{HF}(Q_p) e^{-T(Q_p)} L_{B^2\Sigma}(Q_p=0)|H|R_{A^2\Pi_p}(Q_p=0)e^{T(Q_p)}\Phi_{HF}(Q_p)\rangle}{\partial Q_p} \right|_{Q_p = 0},  \\
  \lambda^{\text{pB}}_p \equiv \left.  \frac{\partial \langle \Phi_{HF}(Q_p) e^{-T(Q_p)} L_{A^2\Pi_p}(Q_p=0)|H|R_{B^2\Sigma}(Q_p=0)e^{T(Q_p)}\Phi_{HF}(Q_p)\rangle}{\partial Q_p} \right|_{Q_p = 0},
\end{eqnarray}
and $sig(\lambda^{\text{Bp}}_p)$  represents the sign of $\lambda^{\text{Bp}}_p$. In the present calculations of linear triatomic molecules, we choose the phase factors for the wave functions so that $\lambda_x = \lambda_y \equiv \lambda$,
i.e.,
%molecular bending modes are symmetric along x and y axis. As long as the phases of wavefunction are properly chosen, we can define $\lambda_x = \lambda_y \equiv \lambda$ for simplicity. 
%Since the Hamiltonian in EOM-CC is not Hermitian, the conjugate matrix elements, such as 
% $V^{\text{LVC}}_{\text{xB}}$ v.s. $V^{\text{LVC}}_{\text{Bx}}$, $h^{\text{SO}}_{\text{AB}}$ v.s. $h^{\text{SO}}_{\text{BA}}$, and $h^{\text{dip}}_{\text{AX}}$ v.s. $h^{\text{dip}}_{\text{XA}}$, are not necessarily the same. Their values were set to be same as the geometric average value in the present paper. In this notation, the linear vibronic coupling (LVC) matrix elements can be written as 
$V^{\text{LVC}}_{\text{xB}} = V^{\text{LVC}}_{\text{Bx}} = \lambda Q_x$ and $V^{\text{LVC}}_{\text{yB}} = V^{\text{LVC}}_{\text{By}} = \lambda Q_y$. 
The inclusion of linear vibronic coupling introduces corrections to the quasidiabatic energies. 
Based on the perturbation theory, the matrix elements in Eq. (5) can be written as
\begin{equation}
  \begin{aligned}
    &{E}^{\text{qd}}_{A^2\Pi_x} = \widetilde{E}^{\text{qd}}_{A^2\Pi_x} + \left.\frac{\lambda^2 }{E^{\text{ad}}_{B^2\Sigma} - \widetilde{E}^{\text{qd}}_{A}}\right|_{Q_x=Q_y=0}   Q_x^2,\\
    &{E}^{\text{qd}}_{A^2\Pi_y} = \widetilde{E}^{\text{qd}}_{A^2\Pi_y} + \left.\frac{\lambda^2 }{E^{\text{ad}}_{B^2\Sigma} - \widetilde{E}^{\text{qd}}_{A}}\right|_{Q_x=Q_y=0}   Q_y^2,\\
    &V^{\text{RT}}_{\text{xy}} = \widetilde{V}^{\text{RT}}_{\text{xy}} + \left.\frac{\lambda^2 }{E^{\text{ad}}_{B^2\Sigma} - \widetilde{E}^{\text{qd}}_{A}}\right|_{Q_x=Q_y=0}   Q_xQ_y,\\
    &{E}^{\text{qd}}_{B^2\Sigma} = E^{\text{ad}}_{B^2\Sigma} - \left.\frac{\lambda^2 }{E^{\text{ad}}_{B^2\Sigma} - \widetilde{E}^{\text{qd}}_{A}}\right|_{Q_x=Q_y=0}   Q_x^2 - \left.\frac{\lambda^2 }{E^{\text{ad}}_{B^2\Sigma} - \widetilde{E}^{\text{qd}}_{A}}\right|_{Q_x=Q_y=0}   Q_y^2,
  \end{aligned}
\end{equation}
in which %The reason we use 
$\widetilde{E}^{\text{qd}}_{A}={E}^{\text{qd}}_{A^2\Pi_x}|_{Q_x=Q_y=0}={E}^{\text{qd}}_{A^2\Pi_y}|_{Q_x=Q_y=0}$.
%is that $\Pi_x$ and $\Pi_y$ share the same energy in the reference structure ($Q_x = Q_y = 0$). 
% The only matrix elements left in the six-state quasi-diabatic Hamiltonian are the spin-orbit coupling (SOC) matrix elements. They are
% \begin{equation}
%   \begin{aligned}
%     & h^{\text{SO}}_{\text{yx}} = i |\langle A^2\Pi_y | H_{SOC} | A^2\Pi_x \rangle| \equiv i H_{AA}^{\text{SO}}\\
%     & \bar{h}^{\text{SO}}_{\text{yx}} = - h^{\text{SO}}_{\text{yx}} = -i H_{AA}^{\text{SO}}\\
%     & h^{\text{SO}}_{\text{Bx}} = -i |\langle B^2\Sigma | H_{SOC} | A^2\Pi_x \rangle| \equiv -i H_{AB}^{\text{SO}}\\
%     & \bar{h}^{\text{SO}}_{\text{Bx}} = h^{\text{SO}}_{\text{Bx}} = -i H_{AB}^{\text{SO}}\\
%     & h^{\text{SO}}_{\text{By}} = |\langle B^2\Sigma | H_{SOC} | A^2\Pi_y \rangle| = H_{AB}^{\text{SO}}\\
%     & \bar{h}^{\text{SO}}_{\text{By}} = -h^{\text{SO}}_{\text{By}} = - H_{AB}^{\text{SO}}\\
%   \end{aligned}
% \end{equation}
% and their complex conjugate parts. 
The spin-orbit matrix elements and transition dipole moments were calculated
using the unrelaxed EOM-CCSD transition density matrices [\onlinecite{Klein08}, \onlinecite{Cheng18a}].
We also used the geometric average values in the parametrization of the quasidiabatic Hamiltonian. 
%The calculated parameters in Eq. (1) together with the transition dipoles are summarized in Table S.\ref{parameters in H table}.

\begin{table}[h]
  \begin{center}
    \caption{{\it{Ab initio}} parameters}
    \label{paraH}
    \begin{tabular}{ccccccc}
      \hline\hline
       & ~ & CaOH/CaOD & ~ & SrOH/SrOD & ~ & YbOH \\
      \hline
      $\lambda$/cm$^{-1}$ & ~ & 121.8/132.7 & ~ & 70.3/79.9 & ~ & 127.3  \\
      $h^{\text{SO}}_\text{AA}$/cm$^{-1}$ & ~ & 33.2 & ~ & 122.2 & ~ & 497.4 \\ 
      $h^{\text{SO}}_\text{AB}$/cm$^{-1}$ & ~ & 26.4 & ~ & 106.1 & ~ & 426.1 \\ 
      $h^{\text{dip}}_{\text{AX}}$/a.u. & ~ & 2.30 & ~ & 2.49 & ~ & 2.32 \\ 
      $h^{\text{dip}}_{\text{BX}}$/a.u. & ~ & 1.85 & ~ & 2.02 & ~ & 1.87 \\
      \hline \hline
    \end{tabular}
  \end{center}
\end{table}

All of the parameters were calculated at the SFX2C-1e-EOMEA-CCSD level with all electrons correlated. The linear vibronic coupling constants, $\lambda$'s, and transition dipoles were obtained using the uncontracted cc-pCVQZ basis sets for CaOH, the cc-pwCVQZ basis sets recontracted for the SFX2C-1e scheme for SrOH, and the cc-pVTZ SFX2C-1e recontracted basis sets for YbOH. The spin-orbit coupling matrix elements were calculated using the corresponding uncontracted basis sets and the SFX2C-1e atomic mean field spin-orbit approach [\onlinecite{zhang2020performance}]. These computed parameters are summarized in Table \ref{paraH}. 

\section{Calculations for the Potential energy surfaces and the discrete variable representation calculations}
The \textit{ab initio} adiabatic potential energy surfaces (PESs) of CaOH, SrOH, and YbOH were obtained from SFX2C-1e [\onlinecite{Dyall2001}, \onlinecite{Liu2009}] EOMEA-CCSD [\onlinecite{Nooijen95}] calculations using the CFOUR program package [\onlinecite{cfour}, \onlinecite{Matthews2020a}]. These calculations were carried out for 2187 grid points for each surface (11 evenly spaced grid points covering the range [$r_{1e}-0.4$ bohr, $r_{1e}+0.6$ bohr] for the M-O bond length, 11 evenly spaced grid points covering the range [$r_{2e}-0.2$ bohr, $r_{2e}+0.3$ bohr] for the H-O bond length, and 18 evenly spaced grid points covering the range [$95^{\circ}$, $180^{\circ}$] for the M-O-H angle, where $r_{1e}$ and $r_{2e}$ refer to equilibrium M-O and O-H bond lengths, respectively). 
The other computational details are documented in Table \ref{PES setup}. 
% PESs of SrOH and YbOH were shifted to the equilibrium structure obtained from quadruple $\zeta$ basis geometry optimization. 
\begin{table}[h]
  \begin{center}
  \caption{Details for the computations of the potential energy surfaces.}
  \label{PES setup}
  \begin{tabular}{cccccccc}
    \hline \hline
    % ~& \multicolumn{3}{l}{CaOH} & ~&\multicolumn{3}{l}{SrOH}\\
     & ~& CaOH & ~& SrOH & ~& YbOH   \\
    \hline
    Basis sets for M &~& cc-pCVQZ  &~& cc-pwCVTZ-X2C &~& cc-pVTZ-X2C  \\
    Basis sets for O/H &~& cc-pVQZ  &~& cc-pVTZ-X2C &~& cc-pVTZ-X2C  \\
    Method    &~& EOMEA-CCSD  &~& \multicolumn{3}{c}{SFX2C-1e-EOMEA-CCSD}  \\
    \# of frozen electrons &~& 12  &~& 20 &~& 48  \\
    $r_{1e}(X^2\Sigma)$ / bohr &~& 3.7415  &~& 3.9794 &~& 3.8651 \\
    $r_{2e}(X^2\Sigma)$ / bohr &~& 1.7916  &~& 1.7976 &~& 1.7961 \\
    $r_{1e}(A^2\Pi)$ / bohr &~& 3.7007  &~& 3.9365 &~& 3.8013 \\
    $r_{2e}(A^2\Pi)$ / bohr &~& 1.7914  &~& 1.7974 &~& 1.7954 \\
    $r_{1e}(B^2\Sigma)$ / bohr &~& 3.7108  &~& 3.9472 &~& 3.8051 \\
    $r_{2e}(B^2\Sigma)$ / bohr &~& 1.7904  &~& 1.7964 &~& 1.7939 \\
    \hline \hline
  \end{tabular}
  \end{center}
\end{table}

The analytical potential energy functions were obtained by fitting \textit{ab initio} energies into a six-order polynomial in terms of three internal coordinate displacements from reference structures $r_1-r_{1e}$, $r_2-r_{2e}$, and $\theta - 180^{\circ}$, i.e., $f(r_1,r_2,\theta) = \sum_{i,j,k}A_{ijk}(1/i!j!k!)(r_1-r_{1e})^i(r_2-r_{2e})^j(\theta-180^{\circ})^k$. 

% The reference structures are not always the same as the equilibrium structures. These parameters are given in Table S.\ref{ref structures}.
% \begin{table}
%   \label{ref structures}
%   \caption{Reference structure parameters used in fitting.}
%   \begin{center}
%     \begin{tabular}{ccccccccc}
%       \hline\hline
%       ~ & ~ & $r_{1e}$/bohr & ~ & $r_{2e}$/bohr & ~ & $\theta_e$/degree \\
%       \hline
%       $X^2\Sigma(\text{CaOH})$ & ~ & 3.7550520102 & ~ & 1.794505623 & ~ & 180.0 \\
%       $A^2\Pi(\text{CaOH})$ & ~ & 3.7550520102 & ~ & 1.794505623 & ~ & 180.0 \\
%       $B^2\Sigma(\text{CaOH})$ & ~ & 3.7550520102 & ~ & 1.794505623 & ~ & 180.0 \\
%       $X^2\Sigma(\text{SrOH})$ & ~ & 3.975590 & ~ & 1.794070 & ~ & 180.0 \\
%       $A^2\Pi(\text{SrOH})$ & ~ & 3.934891 & ~ & 1.793869 & ~ & 180.0 \\
%       $B^2\Sigma(\text{SrOH})$ & ~ & 3.946376 & ~ & 1.793064 & ~ & 180.0 \\
%       \hline \hline
%     \end{tabular}
%   \end{center}
% \end{table}

The discrete variable representation (DVR) [\onlinecite{Colbert1992}] calculations reported here included the four vibrational normal nodes and employed a standard formulation for the kinetic energy operator in the rectangular coordinates. %In the present calculations, the ro-vibrational interactions are omitted. 
The real space basis set was expanded in the normal modes of the $A^2\Pi$ state and consists of 21 evenly spaced points in $[-4.0 Q,4.0 Q]$ for bending modes and the M-O stretching mode, and 28 points in $[-6.8 Q,4.0 Q]$ for O-H stretching for each electronic state, where $Q$ is one unit of the dimensionless normal mode. 

The branching ratio for the transition from $A^2\Pi_{1/2}(000)$ to a vibrational state $\mu$ of the ground electronic state, i.e., $X^2\Sigma_{1/2}(\mu)$ is given by %were calculated from $\nu^3$-weighted formula
\begin{equation}
    b_{\mu}=\frac{(\Delta E)_\mu^3|\Vec{d}_{\mu}|^2}{\sum_{\nu}(\Delta E)_\nu^3|\Vec{d}_{\nu}|^2},
\end{equation}
where $(\Delta E)_\mu$ and $\Vec{d}_{\mu}$ refer to the energy difference and transition dipole moment vector between the $A^2\Pi_{1/2}(000)$ state and the $X^2\Sigma_{1/2}(\mu)$ state. In the present work, the $A^2\Pi_{1/2}(000)$ wave function receives the contributions from six scalar electronic states serving as the electronic basis sets
\begin{eqnarray}
A^2\Pi_{1/2}(000)&=&A^2\Pi_{x,\alpha}(\vec{r},\vec{R})\chi_{A^2\Pi_{x,\alpha}}(\vec{R})
+A^2\Pi_{x,\beta}(\vec{r},\vec{R})\chi_{A^2\Pi_{x,\beta}}(\vec{R}) \nonumber \\
&+& A^2\Pi_{y,\alpha}(\vec{r},\vec{R})\chi_{A^2\Pi_{y,\alpha}}(\vec{R})
+A^2\Pi_{y,\beta}(\vec{r},\vec{R})\chi_{A^2\Pi_{y,\beta}}(\vec{R}) \nonumber \\
&+&B^2\Sigma_{\alpha}(\vec{r},\vec{R})\chi_{B^2\Sigma_{\alpha}}(\vec{R}) 
+B^2\Sigma_{\beta}(\vec{r},\vec{R})\chi_{B^2\Sigma_{\beta}}(\vec{R}),
\end{eqnarray}
while the $X^2\Sigma_{1/2}(\mu)$ wave functions consist of one electronic wave function, e.g,
for the $m_s=1/2$ component,
\begin{eqnarray}
X^2\Sigma_{1/2,1/2}(\mu)&=&X^2\Sigma_\alpha(\vec{r},\vec{R})\chi_{X^2\Sigma,\mu}(\vec{R}),
%X^2\Sigma_{1/2,-1/2}(\mu)&=&X^2\Sigma_\beta(\vec{r},\vec{R})\chi_{X^2\Sigma,\mu}(\vec{R})
\end{eqnarray}
where $\chi$'s represent vibrational wave functions. 
The $\Vec{d}_\mu$ values were calculated within the Franck-Condon (FC) approximation, e.g,
the $\Vec{d}$ value for the transition from $A^2\Pi_{1/2}(000)$ to $X^2\Sigma_{1/2,1/2}(\mu)$ is given by 
%between a vibrational state on $X^2\Sigma$ electronic state and the mixed vibronic state obtained from Eq. (1) can be evaluated by
\begin{eqnarray}
    \Vec{d}_{\mu} 
        &=& \langle A^2\Pi_{x,\alpha}(\vec{r},\vec{R})\chi_{A^2\Pi_{x,\alpha}}(\vec{R}) | {\hat{d}} | X^2\Sigma_\alpha(\vec{r},\vec{R})\chi_{X^2\Sigma,\mu}(\vec{R}) \rangle \nonumber \\
        &+& \langle A^2\Pi_{y,\alpha}(\vec{r},\vec{R})\chi_{A^2\Pi_{y,\alpha}}(\vec{R}) | {\hat{d}} | X^2\Sigma_\alpha(\vec{r},\vec{R})\chi_{X^2\Sigma,\mu}(\vec{R}) \rangle \nonumber \\
        &+& \langle B^2\Sigma_{\alpha}(\vec{r},\vec{R})\chi_{B^2\Sigma_{\alpha}}(\vec{R})  | {\hat{d}} | X^2\Sigma_\alpha(\vec{r},\vec{R})\chi_{X^2\Sigma,\mu}(\vec{R}) \rangle \nonumber \\
        &\approx&
        d_{\text{AX}}^{\text{dip}}\langle \chi_{A^2\Pi_{x,\alpha}} |  \chi_{X^2\Sigma,\mu} \rangle e_x
        +d_{\text{AX}}^{\text{dip}}\langle \chi_{A^2\Pi_{y,\alpha}} |  \chi_{X^2\Sigma,\mu} \rangle e_y
        +d_{\text{BX}}^{\text{dip}}\langle \chi_{B^2\Pi_{\alpha}} |  \chi_{X^2\Sigma,\mu} \rangle e_z,
 %   = \langle \chi_i^{X^2\Sigma} \Phi^{X^2\Sigma} | \Vec{\mu} | \chi^{A^2\Pi_x} \Phi^{A^2\Pi_x} \rangle + \langle \chi_i^{X^2\Sigma} \Phi^{X^2\Sigma} | \Vec{\mu} | \chi^{A^2\Pi_y} \Phi^{A^2\Pi_y} \rangle + \langle \chi_i^{X^2\Sigma} \Phi^{X^2\Sigma} | \Vec{\mu} | \chi^{B^2\Sigma} \Phi^{B^2\Sigma} \rangle,
\end{eqnarray}
in which ${\hat{d}}$ represents the dipole moment operator$, \langle \chi|  \chi\rangle$'s are FC overlap integrals, $e_{x, y, z}$ are unit vectors along the x, y, z axes,
and $d^{\text{dip}}$'s are the electronic transition dipole moment values. 
%where $\Vec{\mu}$ is the dipole moment operator, $\chi$ and $\Phi$ represent the vibrational and electronic wave functions, respectively.  
% weighted by the overlap of vibrational wave function, i.e., $\Vec{d}_{i,f} = (\langle X_i | A_{x,f} \rangle h^{\text{dip}}_{\text{AX}}, \langle X_i | A_{y,f} \rangle h^{\text{dip}}_{\text{AX}}, \langle X_i | B_{f} \rangle h^{\text{dip}}_{\text{BX}})$.

\section{Calculated vibrational energy levels}

\begin{table}
  \begin{center}
  \caption{Calculated relative vibrational energy levels (in cm$^{-1}$) of $X^2\Sigma_{1/2}$, $A^2\Pi_{1/2}$, and $B^2\Sigma_{1/2}$ states of CaOH, CaOD, SrOH, SrOD, and YbOH molecule. The available experimental results from the present work and Refs. [\onlinecite{li1992laser,Presunka1993,Presunka1995,Coxon1992,bernath1984dye,hilborn1983laser,nakagawa1983high,Melville2001,baum2020establishing}] are enclosed in the parentheses.}
  \label{vib ene SI X}
  \begin{tabular}{cclclclclclc}
    \hline \hline
    % ~& \multicolumn{3}{l}{CaOH} & ~&\multicolumn{3}{l}{SrOH}\\
    Vibrational state & ~& CaOH & ~& CaOD & ~& SrOH & ~& SrOD & ~& YbOH  \\
    \hline
    $X^2\Sigma_{1/2}(000)$   &~& 0            &~& 0             &~& 0            &~& 0          &~& 0          \\
    $X^2\Sigma_{1/2}(010)$   &~& 355 (352.6)  &~& 268 (267)     &~& 367 (363.7)  &~& 278 (282)  &~& 322 (319)  \\
    $X^2\Sigma_{1/2}(100)$   &~& 611 (609.0)  &~& 608 (605)     &~& 534 (527.0)  &~& 523 (510)  &~& 532 (528)  \\
    $X^2\Sigma_{1/2}(02^00)$ &~& 695 (688.7)  &~& 522 (519.2)   &~& 702 (703.3)  &~& 537        &~& 631 (627)  \\
    $X^2\Sigma_{1/2}(02^20)$ &~& 718 (713.0)  &~& 539 (536.3)   &~& 736 (735.5)  &~& 556        &~& 654        \\
    $X^2\Sigma_{1/2}(110)$   &~& 954 (956)    &~& 870 (837.8)   &~& 892          &~& 792        &~& 846 (840)  \\
    $X^2\Sigma_{1/2}(03^10)$ &~& 1044 (1018)  &~& 781 (720)     &~& 1048         &~& 802        &~& 952 (947)        \\
    $X^2\Sigma_{1/2}(03^30)$ &~& 1091         &~& 814           &~& 1111 (1111.4)&~& 834        &~& 999        \\
    $X^2\Sigma_{1/2}(200)$   &~& 1214 (1210)  &~& 1211 (1201.1) &~& 1063 (1049.1)&~& 1044       &~& 1062 (1052)\\
    $X^2\Sigma_{1/2}(12^00)$ &~& 1293 (1282.6)&~& 1119          &~& 1223         &~& 1038       &~& 1154 (1144)\\
    $X^2\Sigma_{1/2}(12^20)$ &~& 1310 (1302.0)&~& 1136          &~& 1254         &~& 1059       &~& 1173       \\
    $X^2\Sigma_{1/2}(300)$   &~& 1815         &~& 1815 (1793.8) &~& 1595         &~& 1561       &~& 1600 (1572)\\
    $X^2\Sigma_{1/2}(22^00)$ &~& 1890         &~& 1712          &~& 1740         &~& 1576$^a$   &~& 1680 (1651)       \\
    \hline         
    $A^2\Pi_{1/2}(000)$      &~& 0            &~& 0             &~& 0            &~& 0          &~& 0   \\
    $A^2\Pi_{1/2}(010)$      &~& 346 (345)    &~& 264           &~& 379 (377.8)  &~& 288        &~& 333 (357) \\
    $A^2\Pi_{1/2}(010)$      &~& 362 (360)    &~& 274 (273)     &~& 383 (380.4)  &~& 291        &~& 333 \\
    $A^2\Pi_{1/2}(100)$      &~& 623 (630.7)  &~& 616 (623)     &~& 552 (542.1)  &~& 546        &~& 562 (573) \\
    \hline               
    $B^2\Sigma_{1/2}(000)$   &~& 0            &~& 0             &~& 0            &~& 0          &~& 0 \\
    $B^2\Sigma_{1/2}(010)$   &~& 377          &~& 286           &~& 402 (401)    &~& 306 (311)  &~& 354 \\
    $B^2\Sigma_{1/2}(100)$   &~& 609 (607.4)  &~& 610           &~& 544 (534)    &~& 535 (516)  &~& 559 \\
    $B^2\Sigma_{1/2}(02^00)$ &~& 730          &~& 550           &~& 768 (771)    &~& 589        &~& 687 \\
    $B^2\Sigma_{1/2}(02^20)$ &~& 756          &~& 572           &~& 803 (804)    &~& 609        &~& 712 \\
    $B^2\Sigma_{1/2}(110)$   &~& 971          &~& 818           &~& 936          &~& 833        &~& 898 \\
    $B^2\Sigma_{1/2}(200)$   &~& 1210         &~& 1213          &~& 1084         &~& 1064       &~& 1111 \\
    $B^2\Sigma_{1/2}(12^00)$ &~& 1318         &~& 1148          &~& 1294         &~& 1108       &~& 1228 \\
    $B^2\Sigma_{1/2}(12^20)$ &~& 1338         &~& 1171          &~& 1328         &~& 1128       &~& 1246 \\
    \hline \hline
  \end{tabular}
  \end{center}
  a. Tentative assignment because of the strong anharmonic mixing of normal modes.
\end{table}

The calculated vibrational energy levels of the $X^2\Sigma_{1/2}$, $A^2\Pi_{1/2}$, and $B^2\Sigma_{1/2}$ states of CaOH, CaOD, SrOH, SrOD, and YbOH are documented in Table \ref{vib ene SI X}.

\clearpage

\bibliographystyle{apsrev4-1} % Tell bibtex which bibliography style to use
\bibliography{sroh} 

\clearpage